\newcommand{\CLASSINPUTbottomtextmargin}{1 in}
\documentclass[conference,twoside]{IEEEtran}
\ifCLASSINFOpdf
\else
\fi

\usepackage{amssymb}
\usepackage{amsthm}
\usepackage[cmex10]{amsmath}
\DeclareMathOperator{\E}{\mathbb{E}}

\newcommand {\Define} {\stackrel {\Delta} {=}  }

\newcommand{\mya}{\mathrel{\overset{\makebox[0pt]{{\tiny(a)}}}{=}}}

\newcommand {\pu} {p_{\text{u}}}
\newcommand {\bw} {B_{\text{w}}}
\newcommand{\voi}{\varOmega(i)}

\usepackage{bm}
\usepackage{mathtools}
\usepackage{fixltx2e}
\usepackage{epsfig,makeidx,color}
\usepackage{graphicx,dblfloatfix}
\usepackage{amsbsy}
\usepackage{amssymb}
\usepackage{euscript}
\usepackage{lipsum}
\usepackage[ruled,norelsize]{algorithm2e}
\usepackage{cite}
\usepackage{chngcntr}
\usepackage{lineno}
\usepackage[T1]{fontenc}
\usepackage{microtype}
\usepackage{wasysym}
\usepackage{footnote}

\newtheorem{remark}{\it Remark}

\usepackage[english]{babel}
    \usepackage[font=small,labelsep=space]{caption}
\usepackage{subcaption}    

\IEEEoverridecommandlockouts
\begin{document}
%
\title{Constant Envelope Pilot-Based Low-Complexity CFO Estimation in Massive MU-MIMO Systems}
%
%
%
\author{\IEEEauthorblockN{Sudarshan Mukherjee and Saif Khan Mohammed}
\IEEEauthorblockA{ \thanks{The authors are with the Department of Electrical Engineering, Indian Institute of Technology Delhi (IITD), New delhi, India. Saif Khan Mohammed is also associated with Bharti School of Telecommunication Technology and Management (BSTTM), IIT Delhi. Email: saifkmohammed@gmail.com. This work is supported by EMR funding from the Science and Engineering
Research Board (SERB), Department of Science and Technology (DST),
Government of India.}}
}
\maketitle

\begin{abstract}
In this paper we consider a constant envelope pilot signal based carrier frequency offset (CFO) estimation in massive multiple-input multiple-output (MIMO) systems. The proposed algorithm performs spatial averaging on the periodogram of the received pilots across the base station (BS) antennas. Our study reveals that the proposed algorithm has complexity only linear in $M$ (the number of BS antennas). Further our analysis and numerical simulations also reveal that with fixed number of users and a fixed pilot length, the minimum required transmit pilot power decreases as $\frac{1}{\sqrt{M}}$ with increasing $M$, while maintaining a fixed desired mean squared error (MSE) of CFO estimation.
\end{abstract}


%
\IEEEpeerreviewmaketitle

\section{Introduction}
%
%
%
%
Massive multiple-input multiple-output (MIMO)/large scale antenna systems (LSAS) has been envisaged as one of the key next generation wireless technologies for developing integrated $5$G communication networks with high energy and spectral efficiency and low latency \cite{Andrews, Boccardi}. Massive MIMO is a form of multi-user (MU)-MIMO, where the base station (BS) is equipped with a large antenna array (of the order of hundreds) serving only a few tens of user terminals (UTs) in the same time-frequency resource \cite{Marzetta1}. Increasing the number of BS antennas opens up more available degrees of freedom, resulting in suppression of the multi-user interference (MUI), and providing large array gain \cite{Marzetta2, Ngo1}. These results are however valid only for coherent multi-user detection. Therefore estimation and compensation of carrier frequency offsets (CFOs) of different UTs at the BS is important for practical implementation of massive MIMO systems.

\par Over the past decade, substantial amount of work has already been done on CFO estimation in conventional small scale MIMO systems \cite{Stoica, Ma, Ghogho, Simon, Poor}. These existing results however are not suitable for implementation in massive MIMO systems due to prohibitive increase in their complexity with increasing number of UTs and also with increasing number of BS antennas. Recently in \cite{Larsson2}, the authors studied CFO estimation in massive MIMO systems and have proposed an approximation to the joint maximum likelihood (ML) estimation for the CFOs of all UTs. However this approximation proposed in \cite{Larsson2} requires a multi-dimensional grid search, which has high complexity for large number of UTs (as is the case for multi-user massive MIMO systems). Further \cite{Larsson2} only addresses CFO estimation in frequency-flat fading channel. Subsequently in \cite{gcom2015}, the authors propose a low-complexity correlation-based CFO estimator for massive MIMO frequency-selective channel. However the CFO estimator proposed in \cite{gcom2015} relies on pilot signals having high peak-to-average-power ratio (PAPR). In practice, high PAPR signals tend to get distorted due to channel non-linearity (e.g. non-linear  power amplifiers in the transmitter) and therefore it is desirable that the pilots used for CFO estimation have low-PAPR. To this end, in this paper, we propose a low-complexity technique for CFO estimation, that uses constant envelope pilots.

\par \textsc{Contributions}: In this paper, we propose a spatially averaged periodogram based method for CFO estimation in a massive MU-MIMO frequency-selective channel. The contributions of our work presented in this paper are as follows: (i) we propose a special set of constant envelope (CE) uplink pilots and devise an algorithm for CFO estimation at the BS, using spatially averaged periodogram\footnote[1]{Periodogram of the received pilot signal is computed at each BS antenna, which are then averaged across all BS antennas.}; (ii) analysis of the proposed algorithm shows that the complexity of the proposed CFO estimator increases only linearly with increasing number of BS antennas, $M$. Also the complexity of the proposed estimator increases only linearly with increasing number of UTs ($K$), which is a significant improvement over the exponential complexity of the joint ML estimator in \cite{Larsson2}; (iii) our analysis and numerical simulations also show that with fixed pilot length and fixed $K$, the minimum required pilot transmission power to achieve a fixed desired mean squared error (MSE) of CFO estimation can be reduced approximately by $1.5$ dB with every doubling in $M$ (when $M$ is sufficiently large). Note that this $\frac{1}{\sqrt{M}}$ decrease is also observed in the CFO estimator in \cite{gcom2015}, except the fact that the CFO estimator in \cite{gcom2015} requires high PAPR pilots, while the proposed CFO estimator requires low PAPR CE pilots, which are more practical. Further it is also observed that the CFO estimator proposed in this paper is more energy efficient than the CFO estimator presented in \cite{gcom2015}. [\textbf{{Notations:}} $\E$ denotes the expectation operator and $(.)^{\ast}$ denotes the complex conjugate operator.] 



\section{System Model}

\begin{figure}[t]
\centering
\includegraphics[width= 3.5 in, height= 1 in]{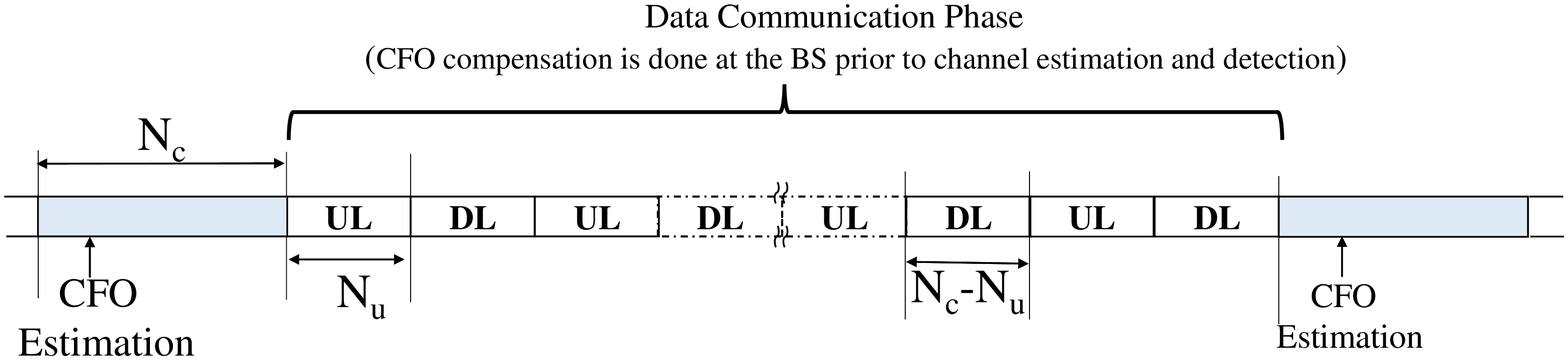}
\caption {The communication strategy: CFO Estimation and Compensation and Data Communication. Here $N_c$ is the duration of coherence interval and the UL slot for data communication is $N_u$ channel uses long. The downlink slot for data communication is therefore of ($N_c - N_u$) channel uses.} 
\label{fig:commstrat}
\end{figure}

Let us consider a single-cell time division duplexed (TDD) massive MIMO BS, equipped with $M$ antennas, serving $K$ single-antenna UTs in the same time-frequency resource. Since the uplink data transmitted by the UTs are coherently detected at the BS, synchronization of the carrier frequency between the BS and the UTs is required for coherent detection. To this end, we propose to perform CFO estimation at the BS in a special coherence slot of $N_c$ channel uses (uplink (UL) plus downlink (DL)) prior to the UL data communication (see Fig.~\ref{fig:commstrat}). In this special slot, the UTs transmit pilots, which are used by the BS to perform CFO estimation. After the special coherence slot for CFO estimation, we have the data communication phase wherein the UTs transmit channel estimation pilots followed by the UL data. During this data transmission phase, the BS performs CFO compensation based on the acquired CFO estimates in the previous special coherence slot, followed by channel estimation and coherent multi-user detection. The special CFO estimation slot can be repeated every few coherence intervals, depending on how fast the CFOs change.

\par In \cite{gcom2015} the proposed CFO estimation technique required temporally separated high PAPR pilot sequences. Since high PAPR sequences are susceptible to channel distortions (e.g. non-linearity of the power amplifiers), thus resulting in degraded performance of the CFO estimator, it is desired to use a low PAPR sequence for CFO estimation in massive MIMO systems. In this paper, we propose constant envelope (CE) low-PAPR pilots, which are not separated in time. Specifically with $K$ UTs, the $k$-th UT transmits the CE pilot $p_k[t] = e^{j\frac{2\pi}{K}(k-1)t}$, $k = 1, 2, \ldots, K$ and $t = 0, 1, 2, \ldots, N-1$, where $N \leq N_c$ is the length of the pilot sequence. Assuming the channel from each UT to the BS to be frequency-selective with $L$ memory taps, the pilot signal received at the $m$-th BS antenna at time $t$ is given by\footnote[2]{A copy of the last $(L-1)$ pilot symbols, i.e., $\{p_k[N-L+1], \cdots, p_k[N-1]\}$ is transmitted before $\{p_k[0], p_k[1], \cdots, p_k[N-1]\}$.}

\begin{IEEEeqnarray}{rCl}
\label{eq:rxpilot}
\nonumber r_m[t] & = & \sqrt{\pu}\sum\limits_{q = 1}^{K}\sum\limits_{l = 0}^{L-1}h_{mq}[l]\, e^{j[\frac{2\pi}{K}(q - 1)(t - l) + \omega_q t]} \, + \, n_m[t]\\
& = & \sqrt{\pu} \sum\limits_{q=1}^{K} H_{mq} \, e^{j[\frac{2\pi}{K}(q - 1) + \omega_q]t} \, + \, n_m[t],
\IEEEeqnarraynumspace
\end{IEEEeqnarray}

\noindent where $H_{mq} \Define \sum\limits_{l = 0}^{L-1} h_{mq}[l] \, e^{-j \frac{2\pi}{K}(q-1)l}$ and $t = 0, 1, 2, \ldots, N-1$. Here $h_{mk}[l] \sim \mathcal{C}\mathcal{N}(0, \sigma_{hkl}^2)$ is the channel gain coefficient from the single-antenna of the $k$-th UT to the $m$-th antenna of the BS at the $l$-th channel tap. $\{\sigma_{hkl}^2\}, \, (l = 0, 1, \ldots, L-1; \, k = 1, 2, \ldots, K)$ model the power delay profile (PDP) of the channel which is perfectly known at the BS. Also $\pu$ is the average power transmitted by each UT. Note that $\omega_q$ is the CFO corresponding to the $q$-th UT and $n_m[t] \sim\mathcal{C}\mathcal{N}(0,\sigma^2)$ is the complex baseband circular symmetric AWGN at the BS. Therefore from \eqref{eq:rxpilot} it is clear that the signal received at the BS is simply a sum of complex sinusoids with additive noise. Specifically the frequency of the sinusoid received from the $k$-th UT is $\frac{2\pi}{K}(k-1) + \omega_k$.


\subsection{Low-complexity CFO Estimation using Spatially Averaged Periodogram}

An intuitive appealing solution to the CFO estimation problem for the above mentioned received pilots is to first obtain an estimate of the frequency of the sinusoid received at the BS from each UT. Since the true frequency of this received sinusoid from the $k$-th UT is $\frac{2\pi}{K}(k-1) + \omega_k$ (see \eqref{eq:rxpilot}), an estimate of the CFO of the $k$-th UT (i.e. $\widehat{\omega}_k$) is simply the difference between the estimate of the received sinusoid frequency from the $k$-th UT and $\frac{2\pi}{K}(k-1)$. It is known that the non-linear least squares (NLS) method is the maximum likelihood joint estimator of the $K$ received frequencies. However it has prohibitive complexity since the objective likelihood function has to be numerically maximized using multi-dimensional grid search in a $K$-dimensional space (see equation (12) in \cite{Larsson2}).

\par An attractive low-complexity alternative to the NLS is the periodogram method, where we simply compute the periodogram of the received signal and choose the location of the largest $K$ peaks as the estimate of the $K$ frequencies received at the BS from the $K$ UTs (this reduces the search space from $K$-dimension to single dimension). However since the received signal power at each BS antenna is expected to be small, we firstly perform spatial averaging of the periodogram computed at each of the $M$ BS antennas. 

\par If all the CFOs are guaranteed to lie in the range $[-\Delta_{\max},\Delta_{\max}]$ (where $\Delta_{\max}$ is the maximum possible CFO for any UT), then the range of the received sinusoid frequency from the $k$-th UT would lie in the interval $[\frac{2\pi}{K}(k-1) - \Delta_{\max}, \frac{2\pi}{K}(k-1) + \Delta_{\max}]$. Since $\Delta_{\max} \ll \frac{\pi}{K}$ in practice\footnote[3]{\label{cforange} For a massive MIMO system with carrier frequency $f_c = 2$ GHz, communication bandwidth $1$ MHz and maximum frequency offset $0.1$ PPM of $f_c$ \cite{Weiss,Ohm}, the maximum CFO is given by $2\pi \times 2 \times 10^{9} \times 0.1 \times 10^{-6}/10^{6} = \frac{\pi}{2500}$. Since in massive MIMO system, the number of UTs $K$ is only of the order of tens, we have $\frac{\pi}{2500} \ll \frac{\pi}{K}$. (A detailed discussion on the range and values of CFOs in massive MIMO system is provided in Remark~1 in \cite{gcom2015}.)}, these intervals are non-overlapping. Hence instead of computing the periodogram over a fine grid in the entire interval $[-\pi,\pi]$, we compute the periodogram only in the interval $[\frac{2\pi}{K}(k-1) - \Delta_{\max}, \frac{2\pi}{K}(k-1) + \Delta_{\max}]$, \, $\forall \, k = 1, 2, \ldots, K$. For the $k$-th UT, the spatially averaged periodogram in the interval $[\frac{2\pi}{K}(k-1) - \Delta_{\max}, \frac{2\pi}{K}(k-1) + \Delta_{\max}]$ is given by

\begin{IEEEeqnarray}{rCl}
\label{eq:contpsd}
\Phi_{k}(\varOmega) & = & \frac{1}{MN}\sum\limits_{m = 1}^{M}\Big | \sum\limits_{t = 0}^{N-1}r_m[t] \, e^{-j[\frac{2\pi}{K}(k - 1) + \varOmega]t}\Big |^2,
\IEEEeqnarraynumspace
\end{IEEEeqnarray}

\noindent where $\varOmega \in [-\Delta_{\text{max}}, \Delta_{\text{max}}]$. The CFO estimate for the $k$-th UT is then given by

\begin{IEEEeqnarray}{rCl}
\label{eq:cfocont}
\widehat{\omega}_{k,0} \Define \arg\max\limits_{ \varOmega \in [-\Delta_{\text{max}}, \Delta_{\text{max}}]} \, \Phi_{k}(\varOmega).
\IEEEeqnarraynumspace
\end{IEEEeqnarray}

\indent In \eqref{eq:cfocont}, the interval $[-\Delta_{\text{max}},\Delta_{\text{max}}]$ is not discrete and therefore in practice, we divide the range of $\varOmega$ into a set of discrete frequencies and compute the periodogram only at those specific frequencies. The proposed set of discrete frequencies is given by $\varXi \Define \{ \varOmega(i) \Define \frac{2\pi}{N^{\alpha}}i \, \big| \, |i| \leq T_0 \}$, where $T_0 \Define \lceil \frac{\Delta_{\text{max}}}{2\pi} N^{\alpha} \rceil$. Since the periodogram has a $\mathcal{O}(1/N)$ resolution, we must have $\alpha > 1$. Therefore the proposed discrete CFO estimator for the $k$-th UT is given by

\begin{IEEEeqnarray}{rCl}
\label{eq:cfoest}
\widehat{\omega}_k = \arg \max\limits_{\varOmega(i) \in \varXi} \, \Phi_k(\varOmega(i)).
\IEEEeqnarraynumspace
\end{IEEEeqnarray}

\par The above proposed algorithm for CFO estimation is summarized in Algorithm~\ref{alg:algo1} at the bottom of the page.

\begin{algorithm}[b]
\caption{CFO Estimation for the $k$-th UT Using Spatially Averaged Periodogram} 
\label{alg:algo1}
\textsc{Input}: $r_m[t]$, $t = 0, 1, \ldots, N-1$ and $m = 1, 2, \ldots, M$;\\
\hspace{1 cm} $\varXi = \{ \varOmega(i) = \frac{2\pi}{N^{\alpha}}i \, \big | \,|i| \leq T_0 \}$, where $T_0 = \lceil \frac{\Delta_{\text{max}}}{2\pi}N^{\alpha}\rceil$.

\par \textsc{Output}: $\widehat{\omega}_k$ (CFO Estimate of the $k$-th UT).\\

\begin{IEEEeqnarray}{rCl}
\label{eq:algoeqn1}
\widehat{\omega}_k = \arg \max\limits_{\varOmega(i) \in \varXi} \, \Phi_k(\varOmega(i)),
\IEEEeqnarraynumspace
\end{IEEEeqnarray}

\noindent where 

\begin{IEEEeqnarray}{rCl}
\label{eq:algoeqn2}
\Phi_k (\varOmega(i)) = \frac{1}{MN}\sum\limits_{m = 1}^{M}\Big | \sum\limits_{t = 0}^{N-1}r_m[t] \, e^{-j[\frac{2\pi}{K}(k - 1) + \varOmega(i)]t}\Big |^2.
\IEEEeqnarraynumspace
\end{IEEEeqnarray}
\end{algorithm}

\begin{figure}[t]
\vspace{-0.3 cm}
\centering
\includegraphics[width= 3.5 in, height= 2.2 in]{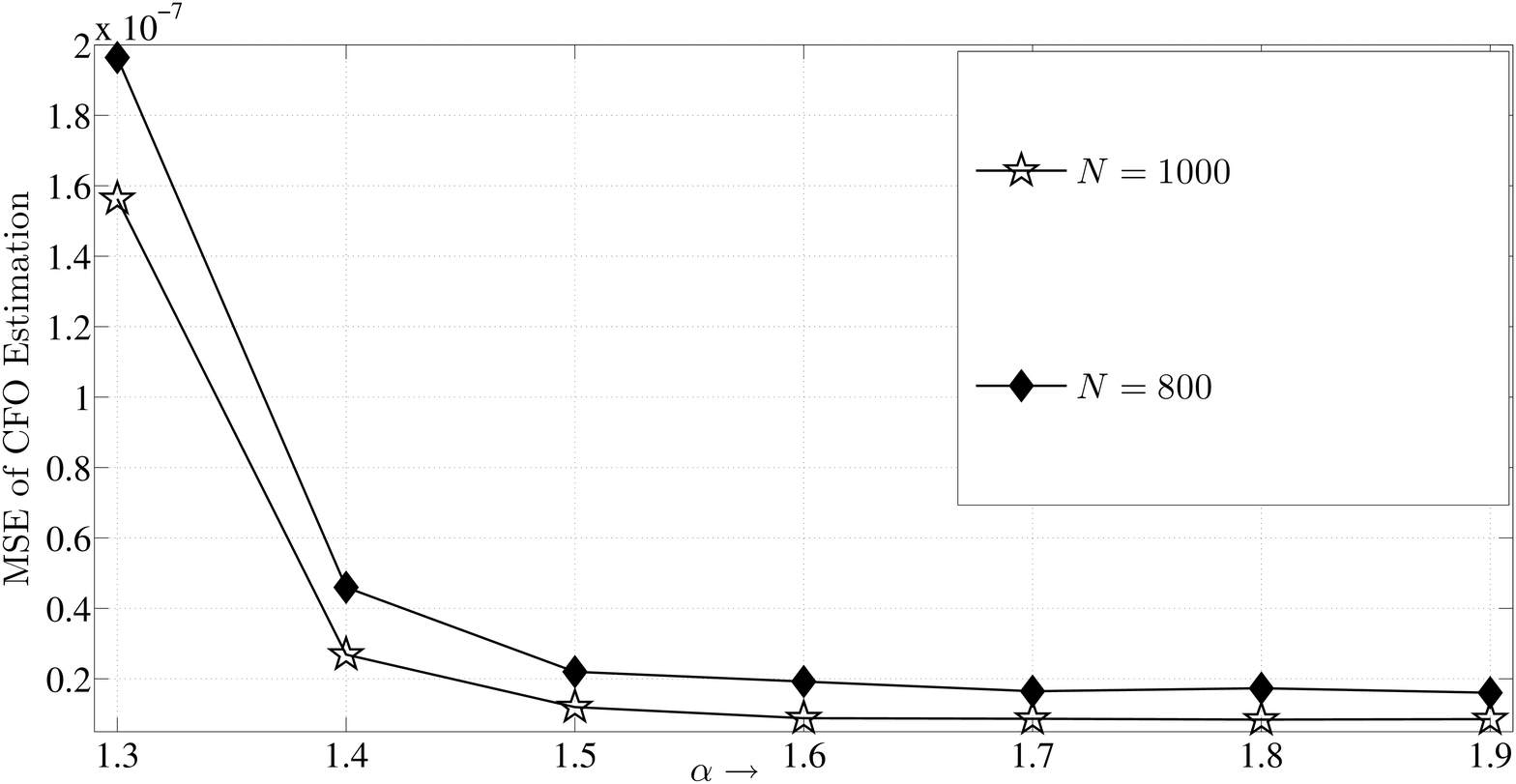}
\caption {Plot of the variation in MSE of CFO estimation with $\alpha$ for a fixed $M = 80$, $K = 10$ and $\gamma = -10$ dB, with $N = 800, 1000$ respectively.} 
\label{fig:alphaopt}
\end{figure}

\begin{remark}
\label{alphaval}
(Optimal value of $\alpha$)
\normalfont
From \eqref{eq:algoeqn1} and \eqref{eq:algoeqn2} it is clear that as $\alpha$ increases, the resolution of the values of $\varOmega(i)$ would also increase, which in turn would improve the accuracy of the CFO estimate. However it is known that even with the NLS method only an $\mathcal{O}(1/N^{3/2})$ accuracy can be achieved \cite{Stoica}. Therefore it is expected that the improvement in the proposed CFO estimate with increasing $\alpha$ would become negligible when $\alpha \geq 3/2$. 

\par The above conclusion is verified from Fig.~\ref{fig:alphaopt}. In Fig.~\ref{fig:alphaopt} we plot the achievable ergodic MSE (numerically computed) versus $\alpha$ for a fixed SNR $\gamma \Define \frac{\pu}{\sigma^2} = -10$ dB, $M = 80$, $K = 10$ and $N = 800, 1000$ respectively. The study shows that in the region $\alpha < 3/2$, the MSE performance improves rapidly with increasing $\alpha$. However when $\alpha \geq 3/2$, the decrease in the MSE becomes negligible irrespective of the value of $N$. Since the complexity would increase with increasing $\alpha$ (see Algorithm~\ref{alg:algo1}), it therefore appears that $\alpha = 3/2$ achieves a good trade-off between the MSE performance and complexity of the proposed CFO estimator.  \hfill \qed 
\end{remark}

\begin{remark}
(Computational Complexity)
\normalfont From \eqref{eq:algoeqn2} we note that the total number of operations required to compute $\Phi_k(\voi)$ is $\mathcal{O}(MN)$. From \eqref{eq:algoeqn1} it is clear that for each user we need to compute $\Phi_k(\voi)$ for $i = -T_0, -T_0+1, \ldots, T_0$, where $T_0 = \lceil \frac{\Delta_{\text{max}}}{2\pi} \, N^{3/2} \rceil$. Hence the total complexity of the proposed CFO estimator is $\mathcal{O}(NKMN^{3/2})$, i.e., a per-channel use complexity of $\mathcal{O}(KMN^{3/2})$. Note that the proposed CFO estimator has a complexity only linear in $M$, i.e., the computational complexity increases linearly with $M$ which is same as the complexity of the low-complexity high-PAPR pilot-based CFO estimator presented in \cite{gcom2015}. Further we also observe that the complexity of the proposed estimator increases only linearly with $K$, which is significantly better compared to the exponential increase in complexity observed in the maximum likelihood joint CFO estimator in \cite{Larsson2}.
\hfill \qed
\end{remark}

\section{Impact of the number of BS antennas on the CFO Estimation Error}

In the following we analyze the impact of the number of BS antennas, $M$, on the CFO estimation error for a fixed number of UTs, $K$ and a fixed length of pilot sequence $N$. We start with studying the expression in the R.H.S. of \eqref{eq:algoeqn2} :

\begin{IEEEeqnarray}{lCl}
\label{eq:psdi0}
\nonumber \frac{1}{\pu} \, \Phi_k(\voi) = \frac{1}{MN\pu}\sum\limits_{m=1}^{M} \Big | \sum\limits_{t = 0}^{N - 1} r_m[t] \, e^{-j [\frac{2\pi}{K}(k-1) + \voi]\, t}\Big |^2\\
 \mya \frac{1}{MN\pu}\sum\limits_{m=1}^{M} \Big | \sum\limits_{t = 0}^{N-1} \Big(\sum\limits_{q=1}^{K}\sqrt{\pu} H_{mq}e^{j\omega_{qki} \, t}\, + \, \widetilde{n}_{mki}[t]\Big)\Big |^2,
\IEEEeqnarraynumspace
\end{IEEEeqnarray}

\noindent where $\omega_{qki} \Define \frac{2\pi}{K}(q-k) + (\omega_q - \voi)$. Here step $(a)$ follows from the expression of $r_m[t]$ in \eqref{eq:rxpilot}. Also $\widetilde{n}_{mki}[t] \Define n_m[t] e^{-j [\frac{2\pi}{K}(k-1) + \voi]\, t} \sim \mathcal{C}\mathcal{N}(0, \sigma^2)$, since $n_m[t]$ is circular symmetric. Expanding the expression in the R.H.S. of \eqref{eq:psdi0}, we further have

\small{
\begin{IEEEeqnarray}{lCl}
\label{eq:psdi1}
\nonumber \frac{1}{\pu} \, \Phi_k(\voi) \mya \frac{1}{M\pu}\sum\limits_{m = 1}^{M} \Big | \sqrt{\pu}\sum\limits_{q=1}^{K}H_{mq}A_k(\omega_q, \voi) \, + \, w_{mki}\Big |^2\\
\nonumber = \underbrace{\frac{1}{M}\sum\limits_{m=1}^{M}\frac{|w_{mki}|^2}{\pu}}_{\Define \, T_1} \, + \, \underbrace{\frac{2}{M}\Re \Big[\sum\limits_{m=1}^{M} \, \frac{w_{mki}^{\ast}}{\sqrt{\pu}}\sum\limits_{q = 1}^{K} H_{mq}A_k(\omega_q, \voi) \Big]}_{\Define \, T_2}\\
 \, + \, \underbrace{\frac{1}{M}\sum\limits_{m=1}^{M}\sum\limits_{q_1 = 1}^{K} \sum\limits_{q_2 = 1}^{K} H_{mq_1} H_{mq_2}^{\ast} A_k(\omega_{q_1}, \voi) \, A_k^{\ast}(\omega_{q_2}, \voi)}_{\Define \, T_3} \, ,
\IEEEeqnarraynumspace
\end{IEEEeqnarray}}\normalsize

\noindent where in step $(a)$ we have $w_{mki} \Define \frac{1}{\sqrt{N}}\sum\limits_{t = 0}^{N-1} \widetilde{n}_{mki}[t] \, \sim \mathcal{C}\mathcal{N}(0,\sigma^2)$ (from the central limit theorem) and $A_k(\omega_q, \voi) \Define \frac{1}{\sqrt{N}}\sum\limits_{t = 0}^{N-1} \, e^{j\omega_{qki} t} \, = \, \frac{1}{\sqrt{N}}\frac{\sin(N\omega_{qki}/2)}{\sin(\omega_{qki}/2)}e^{-j\big(\frac{N-1}{2}\big)\omega_{qki}}$. It is expected that the mean squared error (MSE) of CFO estimation, defined as $\E[(\widehat{\omega}_k - \omega_k)^2]$ would depend on the variance of $\Phi_k(\varOmega(i))$, which is nothing but the sum of the variances of the terms $T_1$, $T_2$ and $T_3$ in \eqref{eq:psdi1} (since $T_i - \E[T_i]$, $i = 1, 2, 3$ are all uncorrelated). The mean and variances of the terms $T_1$, $T_2$ and $T_3$ are summarized in Table~\ref{table:psdterm}. Using the expressions in Table~\ref{table:psdterm} we now study the impact of the number of BS antennas, $M$ on the MSE of CFO estimation.

\begin{savenotes}
\begin{table}[b]
\caption[position=top]{{\textsc{List of Variance and mean of $T_i$, $i = 1, 2, 3$.}}}
\label{table:psdterm}
\centering
\begin{tabular}{| c | c |}
\hline
Component & Variance \vspace{0.01 cm}\\ 
\hline
$T_1$ & $\E[T_1] = \dfrac{1}{\gamma}$, \hspace{0.5 cm} $var(T_1) = \dfrac{1}{M\gamma^2}$ \vspace{0.04 cm} \\
\hline
$T_2$ & $\E[T_2] = 0$,  $var(T_2) = \frac{2}{MN\gamma} \sum\limits_{q = 1}^{K} \, \beta_q \,\frac{\sin^2\left(\frac{N}{2}\omega_{qki}\right)}{\sin^2\left(\frac{1}{2}\omega_{qki}\right)}$, \\
 & where $\beta_q = \sum\limits_{l = 0}^{L-1} \, \sigma_{hql}^2$. \vspace{0.05 cm}\\
\hline
$T_3$ & $\E[T_3] = \frac{1}{N}\sum\limits_{q = 1}^{K} \, \beta_q {\sin^2\left(\dfrac{N}{2}\omega_{qki}\right)}/{\sin^2\left(\dfrac{1}{2}\omega_{qki}\right)}$,\\
 & \hspace{-0.28 cm} $\substack{var(T_3) = \sum\limits_{q_1 = 1}^{K} \sum\limits_{q_2 = 1}^{K} \, \frac{\beta_{q_1} \beta_{q_2}}{MN^2} \frac{\sin^2\big(\frac{N}{2}\omega_{q_{1}ki}\big)\sin^2\big(\frac{N}{2}\omega_{q_{2}ki}\big)}{\sin^2\big(\frac{1}{2}\omega_{q_{1}ki}\big)\sin^2\big(\frac{1}{2}\omega_{q_{2}ki}\big)}}$ \vspace{0.05 cm}\\
\hline
\end{tabular}
\end{table}
\end{savenotes}

\begin{remark}
\label{snrMfixmse}
(Impact of the Number of BS Antennas)
\normalfont From Table~\ref{table:psdterm} note that the variances of all three terms, $T_1$, $T_2$ and $T_3$ decrease with the increasing number of BS antennas, $M$, provided all other system parameters (the pilot length $N$, number of UTs $K$ and transmit SNR $\gamma$) are fixed. From \eqref{eq:psdi1} it is also clear that this reduction in MSE is due to the spatial averaging of the periodogram. At the same time, decreasing $\gamma$ (fixed $M$, $N$, $K$) would however increase the MSE. Therefore we are interested in finding the rate at which $\gamma$ can be reduced with increasing $M$ (fixed $N$ and $K$) while maintaining a fixed desired MSE of the proposed CFO estimator. From Table~\ref{table:psdterm} it is clear that the variance of $T_1$ is proportional to $\frac{1}{M\gamma^2}$ and the variance of $T_2$ is $\propto \frac{1}{M\gamma}$ (fixed $N$ and $K$). Note that the variance of $T_3$ does not depend on $\gamma$. Therefore it appears that with increasing $M$ and (fixed $N$, $K$) and a fixed desired MSE of CFO estimation we can reduce $\gamma$ no faster than $\frac{1}{\sqrt{M}}$, i.e., for every doubling in $M$, $\gamma$ can be reduced approximately by $1.5$ dB when $M$ is sufficiently large. This conjecture has been verified through exhaustive simulations (see Fig.~\ref{fig:snrM}).   \hfill \qed
\end{remark}

\section{Numerical Analysis and Discussions}

In this section through Monte-Carlo simulations, we present a comparative study of the performance of the proposed low-PAPR (constant envelope) pilot based CFO estimator with the high-PAPR pilot-based CFO estimation technique presented in \cite{gcom2015}. For both estimators, we have used the following values for system parameters: operating carrier frequency $f_c = 2$ GHz, communication bandwidth $\bw = 1$ MHz, maximum delay spread of the channel $T_{\text{d}} = 5 \, \mu$s and the channel coherence time $T_c = 1$ ms. The maximum CFO is taken to be $0.1$ PPM of $f_c$ (see also footnote~\ref{cforange}). Therefore we have $\Delta_{\text{max}} = 2\pi \times 0.1 \times 10^{-6} \times \frac{f_c}{\bw} = \frac{\pi}{2500}$. The CFO from each UT is modelled as a uniformly distributed random variable in the interval $[-\Delta_{\text{max}},\Delta_{\text{max}}]$. Also, duration of the coherence interval $N_c = T_c \bw = 1000$, $L = T_{\text{d}} \bw = 5$, number of UTs $K = 10$ and the length of the pilot sequence $N \leq N_c$. Further the PDP is assumed to be the same for each UT and is given by $\sigma_{hkl}^2 = 1/L$ ($l = 0, 1, \ldots, L-1$ and $k = 1, 2, \ldots, K$).

\begin{figure}[t]
\centering
\includegraphics[width= 3.5 in, height= 2.2 in]{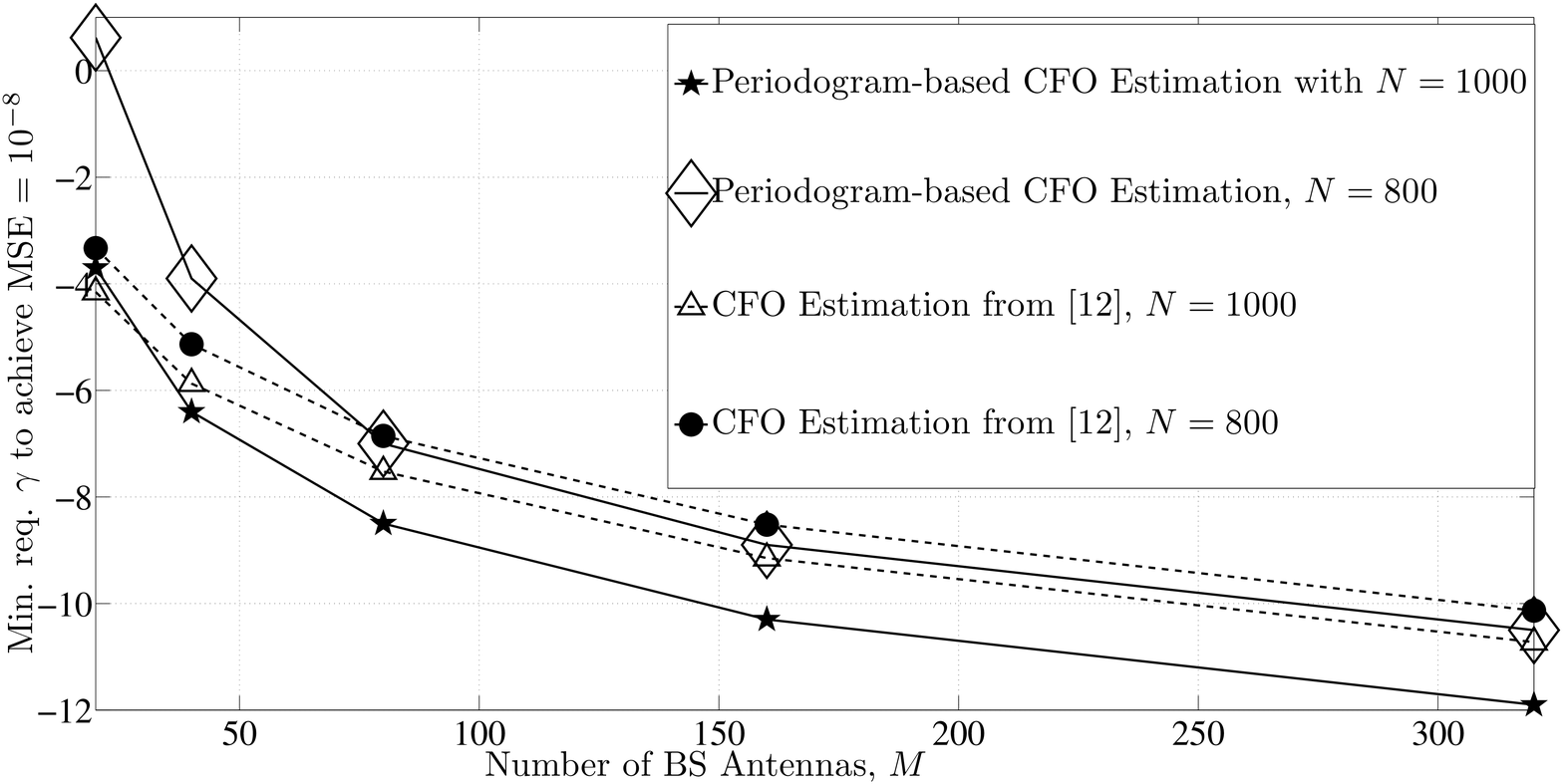}
\caption {Plot of the minimum required SNR to achieve a fixed desired MSE of CFO estimation $\E[(\widehat{\omega}_k - \omega_k)^2] = 10^{-8}$ with increasing number of BS antennas $M$, for the following fixed parameters: $K = 10$, $L = 5$, $N_c = 1000$ and $N = 800, 1000$.} 
\label{fig:snrM}
\end{figure}

\par In Fig.~\ref{fig:snrM}, we plot the variation of the minimum required transmit SNR, $\gamma$ for pilot transmission versus the number of BS antennas, $M$ with fixed $K = 10$ and $N = 800, 1000$ respectively for a fixed desired MSE of CFO estimation $\E[(\widehat{\omega}_k - \omega_k)^2] = 10^{-8}$. From the figure it is clear that for sufficiently large $M$, the required $\gamma$ to achieve MSE $ = 10^{-8}$ decreases roughly by $1.5$ dB with every doubling in $M$ (note the reduction in $\gamma$ from $M = 160$ to $M = 320$ for both $N = 800, 1000$ scenarios). This supports our conclusion on the $\mathcal{O}(\sqrt{M})$ gain in SNR with increasing $M$ (see Remark~\ref{snrMfixmse}).

\par Note that this $\frac{1}{\sqrt{M}}$ decrease in the required SNR for a fixed MSE is also achievable in the CFO estimator proposed in \cite{gcom2015}, except the fact that \cite{gcom2015} uses high-PAPR pilots, which are susceptible to distortion in non-linear channels. In Fig.~\ref{fig:snrM} it is also observed that the proposed CFO estimator requires less SNR than the CFO estimator presented in \cite{gcom2015}, i.e., the proposed CFO estimator is more energy efficient. Also, note that with fixed $M$ and $K$, the required SNR to achieve a fixed desired MSE, decreases with increasing length of pilot sequence ($N$). This is due to the fact that the variances of terms $T_2$ and $T_3$ decrease with increasing $N$ (see Table~\ref{table:psdterm}) and that the resolution of the periodogram is $\mathcal{O}(1/N)$.

\section{Conclusion}
In this paper we propose a low-complexity spatially averaged periodogram-based algorithm for CFO estimation in massive MIMO systems. Contrary to the existing low-complexity algorithm for CFO estimation in \cite{gcom2015}, the proposed CFO estimator uses low-PAPR constant envelope pilots and is more energy efficient. Also the computational complexity of the proposed CFO estimator is only linear with increasing number of BS antennas, $M$, (same as the complexity of the high-PAPR pilot based CFO estimator in \cite{gcom2015}) and also linear in the number of UTs, which is a significant improvement compared to the complexity of the joint ML estimator presented in \cite{Larsson2} (which is exponential in the number of UTs). Study of the performance of the proposed algorithm reveals that for a fixed desired MSE of CFO estimation and for fixed number of UTs and fixed pilot length, the minimum required pilot transmission power of the transmitted pilots decreases as $\frac{1}{\sqrt{M}}$ with increasing $M$.




%

%


\ifCLASSOPTIONcaptionsoff
  \newpage
\fi



%


\bibliographystyle{IEEEtran}
\bibliography{IEEEabrvn,mybibn}

\end{document}